# Dedicated front-end electronics for the next generation of linear collider electromagnetic calorimeter

S.Manen, G. Bohner, J. Lecoq, J. Fleury, C. de la Taille and G. Martin

LPC Clermont-Ferrand, 24 Avenue des Landais , Clermont-Ferrand, France and LAL Orsay, Bâtiment 200, Orsay, France
manen@clermont.in2p3.fr

*Abstract*

This paper describes an R&D electronic program for the next generation of linear collider electromagnetic calorimeter. After a brief presentation of the requirements, a global scheme of the electronics is given. Then, we describe the three different building blocks developed in 0.35μm CMOS technology: an amplifier, a comparator and finally the pipelined ADC.

## I. REQUIREMENTS FOR ECAL

ECAL is a barrel with two end-caps made of sandwich silicon-tungsten structure. It is composed of 40 piled up layers of reading. The basic element for each layer is a diode with 1cm² area.

The first constraint is a great number of channels of measurement. There are 34 Millions channels which represent 3,400m² of silicon. Secondly, we need a great dynamic range, about 15 bits, with accuracy on the data equal to 8 bits. Another important point is the very low consumption required; we have just few mW by channel. And finally the system is relatively slow comparing to LHC. There is one train every 200ms and each one represents 3,000 bunch-crossing with 300ns BX (bunch-crossing) duration.

Presently, the chip is embedded in detector (figure 1). So, the line capacitance is very small and noise requirement about 0.1 MIP (MIP is fixed by silicon wafer depth about 500μm) can be achieved with low preamplifier consumption. Furthermore, the crosstalk is low.

Some drawbacks must be taken into account: the chip is in an electromagnetic shower and some radiations can disrupt good working. Moreover, the chip must be cooled and some compact cooling issues must be studied.

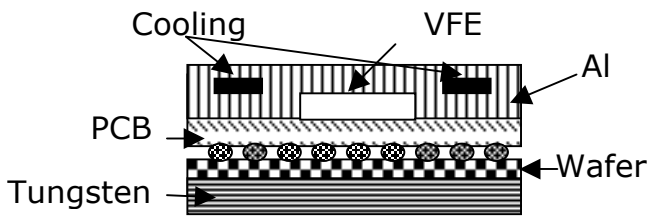

Figure 1: Chip embedded in detector

## II. ELECTRONICS SYNOPTIC

A dedicated integrated circuit must ensure at the same time the analog treatment, the zero suppression and the digitization. So, the front-end chip will be mixed with good analog and digital performances. Furthermore, the technology used must be as cheap as possible. The technology Austria-Microsystems 0.35μm CMOS is chosen. It has two types of transistors: transistors 3.3V for digital blocks and transistors 5V for analog blocks. One possible scheme for electronics of the front-end is represented in figure 2.

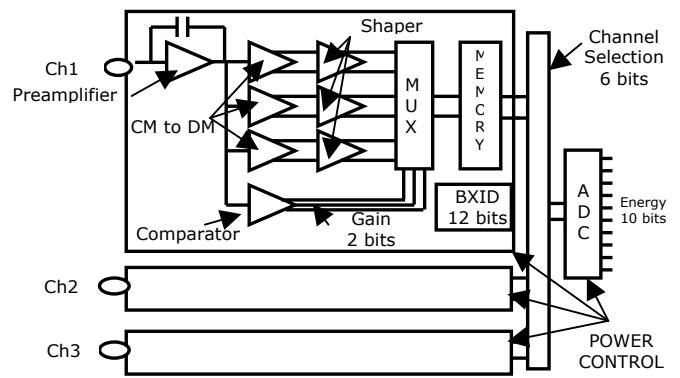

Figure 2: Electronics synoptic

We decide presently to process 36 channels per chip. The signal from the detector, a silicon diode, is shaped. Two alternatives for shaping are studied, a standard method with charge preamplifier and a CRRC² shaper, and a new possibility using a switched integrator. Both methods use multi-gain shaping to handle the 15 bits dynamic range. Those shaped data are then multiplexed and stored in an analog memory because one channel can be touched further times during the same BX (five times max). Finally an ADC provides a data on 10 bits. All this system must be power-controlled to reduce consumption. The analog part works during 1% of time, so a solution (for example) would be to decrease the biasing current by a factor ten when the analog part isn't used. The digital flow is relatively small; we have 34 Millions events per BX coded on 30 bits, so about 1 Gbits per BX.

## III. LOW NOISE PREAMPLIFIER

A very low noise and very low consumption preamplifier is required. This is a standard charge preamplifier developed in LAL Orsay. Serial noise is dominant in our case, so a big transistor PMOS is used in input with W=2000μm, L=0.5μm and $I_{ds}$=200μA to obtain high transconductance. The main characteristics are a power dissipated of 1mW and a noise about 1nV/√Hz.

## IV. TWO ALTERNATIVES OF FILTERING

Two alternatives of shaping are studied, a CRRC² shaper and a switched integrator.

### A. CRRC² Shaper

The CRRC² shaper is a standard alternative of filtering.

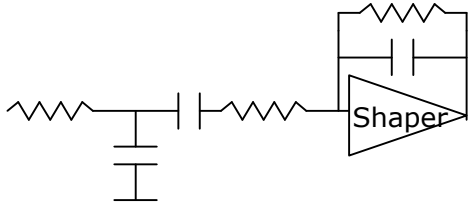

Figure 3: CRRC² shaper

With a serial noise about 1nV/√Hz, we obtain an equivalent noise charge equal to 1,400 electrons. This system is very well known but the resistors take large area on the chip and the reset is slow, so we have pile-up.

### B. Switched integrator

The switched integrator is a non standard method of filtering. The figure 3 presents this type of filtering.

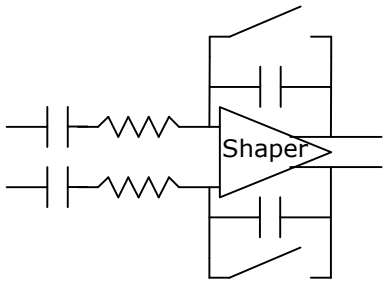

Figure 4: Switched integrator

The shaper is a fully differential integrator. The equivalent noise charge is equal to 1,600 electrons. This system presents a lot of advantages : the switch is easier integrated than large resistors and furthermore, the reset is fast so there isn't pile-up. However, the switches must be ordered with a syncronous signal.

## V. CMOS AMPLIFIER

We have developed an amplifier CMOS. Two prototypes have been produced and tested in 0.8μm BiCMOS and 0.35μm CMOS Austria Microsystems technologies.

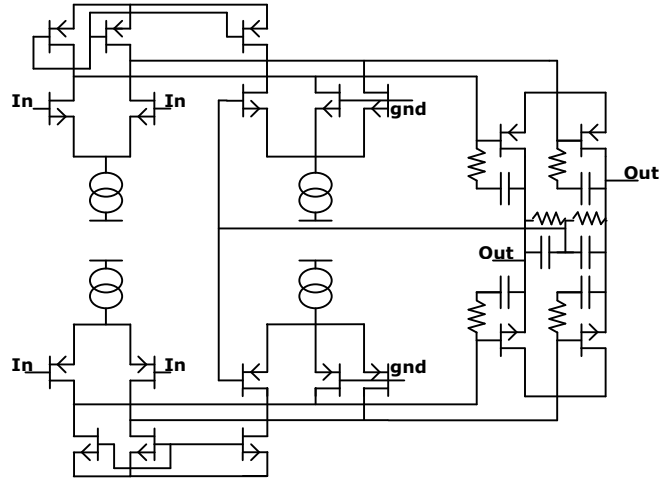

Figure 5: Simplified scheme of the fully differential input output rail to rail amplifier.

This amplifier is a concatenation of two known schematics: the fully differential output amplifier with resistive common mode feedback and the rail to rail input output amplifier. The first concept allows a full symetrical design around a well control of the common mode operating point. The second helps us to make the input stage gain independent of the signal amplitude and also to achieve a very good CMRR.

The main characteristics of this amplifier are a differential power supply of 2.5V with consumption about 300μA and a gain bandwidth product about 50MHz. The offset measured is 1mV, so we have a good matching.

On the figure 6, test result of the switched integrator is shown. There is a perfect integration with a reset time about 20ns for 3V maximum amplitude input.

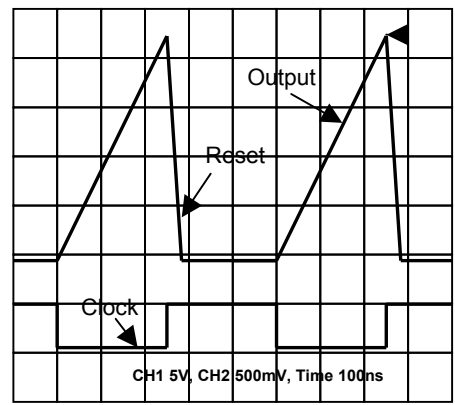

Figure 6: Switched integrator tests results

## VI. 10 BITS PIPELINED ADC

The third block is the 10 bits pipelined ADC. It is composed of a comparator and a gain 2 amplifier with capacitive feedback. Two prototypes in technologies 0.35μm have been developed and the total consumption of this block is 20mW. The principle of conversion is based on 10 stages to convert 10 bits. The algorithm defines two possible cases:

- If $V_{in} < V_{ref}$ then bit=1 and $V_{in+1} = (V_{in} - V_{ref})*2$
- If $V_{in} > V_{ref}$ then bit=0 and $V_{in+1} = V_{in}*2$

So to process this algorithm, we need an accurate gain 2 amplifier and a comparator.

### A. Gain 2 amplifier

The amplifier used is the same than shaper one. To obtain an accuracy of 1/1000 on gain 2, we use capacitors because of better matching than resistors. The drawing of the 300fF capacitance represents an area of 18.5*18.5μm². The capacitance layout is designed with a lot of care to match parasitic capacitors.

The tests results are a gain with accuracy better than 1% and an average offset -4mV with a sigma ±4mV.

### B. Comparator

The second block used in the ADC is the comparator.

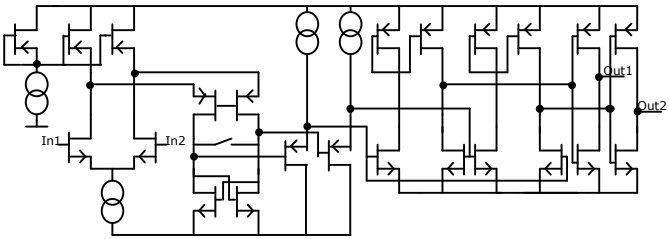

Figure 7: CMOS Comparator

This is a latched comparator with a differential folded cascode input pairs. Two different prototypes have been designed in 0.8μm BiCMOS and 0.35μm CMOS technologies.

The consumption of this element is 100μA under ±2.5V with sensitivity of 300μV. The offset measured of this first version is too high, about 11mV due to parasitic capacitance non equal on each side of the differential pairs.

## VII. ADC PROTOTYPES

### A. 10bits ADC in 0.35μm CSI

The first prototype has been developed in 0.35μm CSI. The tests results give us only 5 bits due to problems of instability, but the functionality is verified. The main characteristics are a differential power supply of ±2.5V; the dissipation is 20mW with differential input range of ±1V and a clock frequency of 5MHz.

### B. 10bits ADC in 0.35μm C35B4

This new technology permits to draw bigger resistors occupying fewer areas due to polysilicon high resistivity layer. Several modifications have been done on this version. The first one is to stabilize the amplifier. We have just increased the RC in output of the amplifier. Second, a new design with a more compact drawing of capacitors has been tested to improve the gain 2 accuracy. Third one is to improve the offset comparator; for this, the parasitic capacitances on each side of the differential pairs have been matched. Finally, power supply decoupling has been improved.

With all those improvements, the parasitic simulation gives us a correct working on 10 bits which could be verified with tests at the end of October.

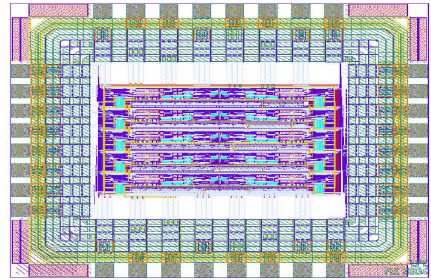

Figure 8: Die picture of the pipelined 10 bits ADC

## VIII. SUMMARY

This paper has described one possible scheme for the front end electronics dedicated to the electromagnetic calorimeter of the linear collider. Two alternatives of filtering have been presented, one with a CRRC² shaper and the other one with a switched integrator. Presently, some blocks exist; the charge preamplifier developed by LAL in Orsay, the switched integrator and a very low consumption CMOS comparator.

Concerning ADC, after the tests of the second version, a lot of improvements could be done. For example, a conversion of 1.5 bit per stage with a digital correction could be studied because this is the best compromise between consumption and resolution. Last but not least, to reduce consumption, only one current master for all amplifiers and comparators is a possibility which implies some layout problems.

Finally, the pulsed power supply for analog part must be studied in details.